\documentclass[11pt,a4paper,twoside]{article}
\usepackage{amssymb}%,textcomp,mathpple}
 \usepackage{latexsym}
\evensidemargin\oddsidemargin

\long\def\comment#1{{}}   %%% to comment out some parts
\def \pref#1{{\mathbf{pref}({#1})}}

 \def \card {\mathrm{card}\,}
\newcommand{\x}{{\mathbf x}}
\newcommand{\y}{{\mathbf y}}

\def \bbbn {\mathrm{I\!N}} %%% natural numbers {\Bbb{N}}
\def \XF  {X^{(f)}}
\def \raisedot {\raisebox{.5ex}{.}}
\def \raisecomma {\raisebox{.5ex}{,}}

\newtheorem{theo}{Theorem}
\newtheorem{lem}[theo]{Lemma}

\newtheorem{open}[theo]{Open Question}
\newtheorem{ex}[theo]{Example}

\newenvironment{proof}{\textbf{Proof.\,}}{\hfill\textbf{\itshape
q.e.d.}\par}

\begin{document}
\title{\bf Randomness  Relative to Cantor Expansions}

\author{\bfseries Cristian S. Calude\\Department of Computer
Science\\The University of Auckland\\Private Bag 92019\\Auckland, New
Zealand\\
{\tt
cristian@cs.auckland.ac.nz}\and {\bf Ludwig Staiger}\\
Martin-Luther-Universit\"at
Halle-Wittenberg\\ Institut f\"ur Informatik\\ {D\,-\,06099} Halle,
Germany\\
{\tt
staiger@informatik.uni-halle.de}
\and
{\bf Karl  Svozil}\\
Institut f\"ur Theoretische Physik\\
 University of Technology Vienna\\
 Wiedner Hauptstra\ss e 8-10/136\\
 A-1040 Vienna, Austria\\{\tt svozil@tph.tuwien.ac.at}
} \date{\today}
\maketitle
\noindent
\parindent0pt \thispagestyle{empty}

\begin{abstract}
Imagine a sequence in which the first letter comes from a binary alphabet,
the second letter can be chosen on an alphabet with 10 elements, the
third letter  can be chosen on an alphabet with 3 elements and so on.
When such a sequence can be called {\it random}? In this paper we offer
a solution to the above question using the approach to randomness proposed
by
Algorithmic Information Theory.
\end{abstract}

\section{Varying Alphabets and the Cantor Expansion}

Algorithmic Information Theory (see \cite{ch8,gregc2,Ca})
deals with random sequences over a finite (not necessarily binary)
alphabet. A real number is random if its binary expansion is
a binary random sequence; the choice of base is irrelevant
(see \cite{Ca} for various proofs).

Instead of working with a fixed alphabet we can imagine that the letters of
a sequence are taken from a {\it fixed}
sequence of alphabets.  This construction was introduced by Cantor
as a generalization of the $b$--ary
expansion of reals.  More precisely, let
\[b_1, b_2, \ldots b_n, \ldots\]
be a fixed infinite sequence of positive integers greater than 1. Using a
point we form the finite or infinite sequence
\begin{equation}
\label{cantor}
0.x_{1}x_{2} \ldots
\end{equation}
such that  $0 \le x_n \le b_n-1$, for all $n\ge 1$. Consider the set of
rationals
\begin{equation}
\label{eq1}
s_1 = \frac{x_1}{b_1}  \raisebox{.5ex}{,}  s_2 = \frac{x_1}{b_1} +
\frac{x_2}{b_1b_2}
\raisebox{.5ex}{,}  \cdots \raisebox{.5ex}{,}  s_n =
s_{n-1} + \frac{x_n}{b_1 b_2 \cdots b_n}\raisebox{.5ex}{,} \cdots
\end{equation}

The above sum is bounded from above by 1,
\[0 \le s_n \le \sum_{i=1}^n \frac{b_i - 1}{b_1 b_2 \ldots b_i} = 1 -
\frac{1}{b_1 b_2 \ldots b_n} < 1,\]
so there is a unique real number $\alpha$ that is the least upper bound of
all partial sums (\ref{eq1}). The sequence (\ref{cantor}) is called the {\it
Cantor  expansion}
of the real $\alpha \in [0,1]$.

If $x_n = b_n -1$, for all $n \ge 1$, then $s_n = 1 -
1/(b_1 b_2 \ldots b_n),$ so $\alpha =1$.
If $b_n = b$, for all $n\ge 1$, then the Cantor expansion becomes the
classical $b$--ary expansion. If $x_n =1$ and $b_n = n+1$, for all $n \ge
1$, then $\alpha = e$.

The genuine strength of the Cantor expansion unfolds
when various choices and  interactions on different scales are
considered.

The main result regarding Cantor expansions is:

\begin{theo}
Fix an infinite sequence of scales $b_1, b_2, \ldots$.  Assume that we
exclude Cantor expansions in which starting from some place after the point
all the consecutive digits are $x_n = b_n -1$. Then, every real number
$\alpha\in [0,1]$ has  a unique Cantor expansion (relative to $b_1, b_2,
\ldots$)
and its digits are determined by the following relations:
\[ \rho_1 = \alpha, x_1 = \lfloor b_1 \rho_1\rfloor, \rho_{n+1} = b_n \rho_n
- x_n, x_{n+1} = \lfloor b_{n+1} \rho_{n+1}\rfloor.\]
\end{theo}

Consequently, if we exclude Cantor expansions in which starting from some
place after the point all the consecutive digits are $x_n = b_n -1$. then
given $\alpha \in [0,1]$ there is a unique sequence $\x^{\alpha}\in \XF$
whose Cantor expansion is exactly $\alpha$. If $\x \in \XF$, then we denote
by $\alpha^{\x}$ the real whose Cantor digits are given by the sequence
$\x$, hence $\x^{\alpha^{\x}} = \x$ and $ \alpha^{\x^{\alpha}}
 = \alpha$.

For more
details regarding the Cantor expansion see \cite{hw, drobot}.

\section{Examples}

First, following \cite{drobot}  we consider the British system in which
length can be measured in miles, furlongs, chains, yards, feet, hands,
inches, lines. These scales relate in the following way:  1 mile = 8
furlongs = $8 \cdot 10$ chains =
$8 \cdot 10 \cdot 22$ yards =  $8 \cdot 10 \cdot 22 \cdot 3$ feet = $8 \cdot
10 \cdot 22 \cdot 3
\cdot 4$ hands = $8 \cdot 10 \cdot 22 \cdot 3
\cdot 4 \cdot 3$ inches  =  $8 \cdot 10 \cdot 22 \cdot 3
\cdot 4 \cdot 3 \cdot 12$ lines. Hence,  the sequence of  scales starts with
$b_1 = 10, b_2 = 8, b_3 = 10, b_4 = 22,  b_5 = 3, b_6 = 4, b_7 =3, b_8 = 12$
and can be continued ad infinitum.  For example,  the number $0.963
(11)232(10)00\cdots 0 \cdots$
represents a length of 9 miles, 6 furlongs, 3 chains, 11 yards, 2 feet, 3
hands,  2 inches and 10 lines.

For our second example we consider a ball in gravitational fall
impinging onto a  board of nails with different numbers
$ b_n+1$ of nails
at different horizontal levels
(here, $n$ stands for the $n$th horizontal level
and $b_n$ is the basis corresponding to the position $n$).
Let us assume that the layers are ``sufficiently far apart''
(and that there are periodic boundary conditions realizable by
elastic mirrors).
Then, depending on which one of the $b_n$ openings the ball takes,
one identifies the associated number
(counted from $0$ to $b_n-1$) with
the $n$th position  $x_n \in \{0,\ldots ,b_n-1\}$ after the point.
The resulting sequence leads to the real number  whose Cantor expansion
is $0.x_1x_2\cdots x_n \cdots $.

As a third example we consider a quantum correspondent to the  board of
nails
harnessing irreducible complementarity and the randomness  in the outcome of
measurements on single  particles.
Take a quantized system with at least two  complementary
observables
$\hat{A},\hat{B}$,
each one associated with $N$ different outcomes $a_i,b_j$, $i,j \in
\{0,\ldots ,N-1\}$, respectively.
Notice that, in principle, $N$ could be a large (but finite) number.
Suppose further that $\hat{A},\hat{B}$ are ``maximally''  complementary
in the sense that measurement of $\hat{A}$ totally randomizes the outcome
of $\hat{B}$ and {\it vice versa}
(this should not be confused with optimal mutually unbiased measurements
\cite{WooFie}).

A real number $0.x_1x_2\cdots x_n \cdots $ in the Cantor expansion
can be constructed from successive  measurements of $\hat{A}$
and $\hat{B}$ as follows.
Since all bases $b_n$ used for the Cantor
expansion are assumed to be bounded, choose $N$ to be
the least common multiple of all bases $b_n$.
Then partition the $N$ outcomes into even partitions,
one per different base,
containing as many elements as are required for
associating different elements of the $n$th partition
with  numbers from the set  $\{0,\ldots ,b_n-1\}$.
Then, by measuring
$$\hat{A},\hat{B},\hat{A},\hat{B},\hat{A},\hat{B},\ldots$$
successively, the $n$th position  $x_n \in \{0,\ldots ,b_n-1\}$
can be identified
with the number associated with the element of the partition which
contains the measurement outcome.

As an example, consider the Cantor expansion of a number in the bases
 2, 6, and 9. As the  least common multiple is 18, we choose two
observables with 18 different outcomes; e.g., angular momentum
components in two perpendicular directions of a particle of total
angular
momentum  ${9\over 2} \hbar$ with outcomes in  (units are in $\hbar$)
$$\left\{ -{9\over 2} \raisebox{.4ex}{,}-4,-{7\over 2}
\raisebox{.4ex}{,}
\raisebox{.4ex}{.}\raisebox{.4ex}{.}\raisebox{.4ex}{.,} +{7\over 2} \raisebox{.4ex}{,} +4,+{9\over
2}\right\} \raisebox{.4ex}{.}$$
Associate with the outcomes the set
$\{0,1,2,\ldots ,17\}$
and form the even partitions
\begin{eqnarray}
&\{\{0,1,2,3,4,5,6,7,8\},\{9,10,11,12,13,14,15,16,17\}\},&\nonumber  \\
&\{\{0,1,2\},\{3,4,5\},\{6,7,8\},\{9,10,11\},\{12,13,14\},\{15,16,17\}\},&\nonumber  \\
&\{\{0,1\},\{2,3\},\{4,5\},\{6,7\},\{8,9\},\{10,11\},\{12,13\},\{14,15\},\{1
6,17\}\},&\nonumber
\end{eqnarray}
(or any partition obtained by permutating the elements of $\{0,1,2,\ldots
,17\}$)
associated with the bases  2, 6, and 9, respectively.

Then, upon successive measurements of angular momentum
components in the two perpendicular directions,
the outcomes  are translated into random digits in the bases  2, 6, and 9,
accordingly.

As the above quantum example may appear   ``cooked up",  since  the
coding is based on a uniform radix $N$ expansion, one
might consider successive measurements of the location and the velocity of
a single particle. In such a case, the value $x_n$ is obtained by
associating with it
the click in a particular detector (or a range thereof) associated with
spatial or momentum measurements. Any such arrangements are not very
different in principle, since every measurement of a quantized system
corresponds to registering a discrete event associated with a detector
click \cite{sum-4}.
%The genuine strength of the Cantor expansion unfolds when we consider
% varying choices and varying interactions on different scale levels.

\if01

This is a generalisation of self-similarity with intrinsic scale dependence.
Geometric objects of this type might not scale in a self-similar manner
but could be codable by a Cantor expansion.

Consider a generalized Cantor set obtained, for example,
by cutting out in the $n$th construction step $1/(n+1)$th
of each of the remaining segments (starting from the real interval $[0,1]$
at $n=1$)
at a random position of $n+1$ positions of equal length \cite{nn}.
In Figure \ref{20303-cc-f1}, the construction process is depicted.
\begin{figure}
\begin{center}
%TexCad Options
%\grade{\off}
%\emlines{\off}
%\beziermacro{\on}
%\reduce{\on}
%\snapping{\off}
%\quality{2.00}
%\graddiff{0.01}
%\snapasp{1}
%\zoom{1.00}
\unitlength 0.6mm
\linethickness{1.5pt}
\begin{picture}(90.00,40.00)
\put(0.00,40.00){\line(1,0){90.00}}
\put(90.00,40.00){\line(0,0){0.00}}
\put(0.00,30.00){\line(1,0){30.00}}
\put(60.00,30.00){\line(1,0){30.00}}
\put(0.00,20.00){\line(1,0){7.67}}
\put(16.00,20.00){\line(1,0){14.00}}
\put(60.00,20.00){\line(1,0){11.00}}
\put(79.00,20.00){\line(1,0){11.00}}
\put(0.00,10.00){\line(1,0){1.67}}
\put(4.33,10.00){\line(1,0){3.33}}
\put(17.67,10.00){\line(1,0){12.33}}
\put(59.67,10.00){\line(1,0){6.33}}
\put(68.33,10.00){\line(1,0){2.67}}
\put(78.67,10.00){\line(1,0){9.67}}
\put(0.00,0.00){\line(1,0){0.33}}
\put(0.67,0.00){\line(1,0){1.00}}
\put(4.33,0.00){\line(1,0){0.33}}
\put(5.00,0.00){\line(1,0){2.33}}
\put(17.67,0.00){\line(1,0){6.00}}
\put(24.67,0.00){\line(1,0){5.33}}
\put(60.00,0.00){\line(1,0){2.33}}
\put(63.00,0.00){\line(1,0){2.67}}
\put(68.33,0.00){\line(1,0){0.33}}
\put(69.00,0.00){\line(1,0){1.67}}
\put(78.67,0.00){\line(1,0){2.67}}
\put(83.00,0.00){\line(1,0){5.00}}
\end{picture}
\caption{\label{20303-cc-f1}First construction steps of a generalized
Cantor set.}
\end{center}
\end{figure}

Another example is the generalized Koch curve obtained by inserting in the
$n$th
construction step $n+1$ scaled down copies of the object obtained in the
$n$s
construction step at random positions.
A different variation of the Koch curve is obtained if different objects
(as compared to previous construction steps) are inserted.
Any one of the above examples can be efficiently coded by
Cantor expansions of random reals.
For an efficient encoding,
associate with every construction step a place in the expansion.
Then, the basis chosen for this particular place in the expansion
should be identified with the number of different segments in that
construction step.
For example, in the generalized Cantor set discussed above, there are $n+1$
segments
at the $n$th construction level; therefore, the basis chosen for the
$n$th position
should be $n+1$. (This linear dependence is only an example, and much more
general functions for the bases are possible.)

It is not too speculative to assume that this might reflect the physical
property
of different object formations at different (e.g., length or time) scales,
which might
be caused by nonsimilar interactions at different scales.
\fi

\section{Notation and Basic Results}

We consider $\bbbn$ to be the set of non-negative integers. The
cardinality of the set $A$ is denoted by $\card(A)$.  The base 2 logarithm
is denoted by
$\log$.

If $X$ is a  set, then $X^*$ denotes the free monoid (under concatenation)
generated by $X$ with $e$ standing for the empty string. The length of a
string $w \in X^*$ is denoted by $|w|$. We consider the space $X^\omega$ of
infinite sequences ($\omega$-words)
over $X$. If $\x = x_{1}x_{2}\ldots x_{n}\ldots \in X^{\omega}$, then
$\x(n)=
x_{1}x_{2}\ldots x_{n} $ is the prefix of length $n$ of $\x$. Strings and
sequences will be denoted respectively by $x, u,v,v,w, \ldots$ and   $\x,
\y,
\ldots$. For $w, v\in X^*$ and $\x \in X^\omega$ let  $wv, w\x$ be the
concatenation between $w$ and $v, \x$, respectively.
%The concatenation product extends naturally to subsets $W \subseteq X^*$
%(languages) and $B \subseteq X^*\cup X^\omega$.

By ``$\sqsubseteq$'' we denote the prefix relation between strings:
$w\sqsubseteq v$ if there is a $v'$ such that $w v'= v$.  The relation
``$\sqsubset$'' is similarly defined for $w\in X^{*}$ and $\x \in
X^{\omega}$:
$w\sqsubset\x$ if there is a sequence $\x'$ such that $w\x'= \x$. The sets
$\pref{\x} =\{w : w\in X^*,  w\sqsubset\x\}$ and $\pref{B}=\bigcup_{\x\in
  B}\pref{\x}$ are the languages of prefixes of $\x \in X^{\omega}$ and $B
\subseteq X^{\omega} $, respectively. Finally, $wX^{\omega} = \{\x \in
X^{\omega} : w\in\pref{\x} \}.$ The sets $(wX^{\omega})_{w\in X^*}$ define
the
natural topology on $X^{\omega}$.

Assume now that $X$ is finite and has $r$ elements. The unbiased discrete
measure on $X$ is the probabilistic measure $h (A) =
\card(A)/r$, for every subset of $X$. It induces the product measure $\mu$
defined on all Borel subsets of $X^{\omega}$. This measure coincides with
the
Lebesgue measure on the unit interval, it is computable and $\mu(w
X^{\omega})
= r^{-|w|}$, for every $w \in X^{*}$. For more details see
\cite{mar1,mar2,Ca}.

In dealing with  Cantor expansions we  assume that the sequence of bases
$b_1, b_2, \ldots b_n, \ldots$ is computable, i.e.  given by a computable
function  $f:\bbbn\to \bbbn\setminus\{0,1\}$. Let  $X_i=\{0,\dots,
f(i)-1\}$, for
$i\ge 2$,  and define the space $$X^{(f)} = \prod_{i=1}^\infty X_i\subseteq
\bbbn^\omega\ .$$
The set
$$\pref {X^{(f)}} = \{ w : w = w_1 w_2 \ldots w_n, w_i \in X_i, 1 \le i \le
n\}$$
plays for $X^{(f)}$ the role played by $X^*$ for $X^{\omega}$.

Prefixes of a sequence $\x \in X^{(f)}$  are defined in a natural way and
the set of all (admissible) prefixes  will be denoted by $\pref{\x}$.  As we
will report any coding to binary, the length of $w= w_1 w_2 \ldots w_n\in
\pref {X^{(f)}}$ is
$\parallel w \parallel = \log (\prod_{i=1}^{n} f(i)); |w| =n$. In $X^{(f)}$
the topology is induced by the
sets $[w]_{f} = \{\x \in
X^{f} : w\in\pref{\x} \}$ and the corresponding measure is defined by
$$\mu([w]_{f})
= \prod_{i=1}^{|w|} (f(i)^{-1}),$$ for every $w \in \pref{X^{f}}$.
An open set is of the form $[A]_f = \{\x : \exists n (\x(n)\in A)\}$, for some set
$A\subseteq
\pref{\XF}$. The open set $[A]_f$ is computably enumerable if $A$ is
computably enumerable.
Only the equivalence between the notions of Cantor--randomness
and
weakly Chaitin-Cantor--randomness will be proven

The following two lemmas will be useful:

\begin{lem}
\label{binapprox}
Let $0 \le a < 2^{m}$ and let $\alpha, \beta$ be two reals in the interval
$[a\cdot 2^{-m}, (a+1)\cdot 2^{-m}]$. Then, the first $m$ bits of $\alpha$
and $\beta$ coincide, i.e., if $\alpha = \sum_{i=1}^{\infty} x_{i}2^{-i}$
and $\beta = \sum_{i=1}^{\infty} y_{i}2^{-i}$, then $x_{i}=y_{i}$, for all
$i =1,2, \ldots ,m$.
\end{lem}

\begin{lem}
\label{cantorapprox}
Let $b_{1}, b_{2}, \ldots$ be an infinite sequence of scales
and $a= j/(b_{1}b_{2}\ldots b_{m})\in [0,1]. $  Let $\alpha, \beta$ be two reals in the interval
$[a, a + 1/(b_{1}b_{2}\ldots b_{m})]$. Then, the first $m$ digits of the
Cantor expansions (relative to $b_{1}, b_{2}, \ldots$) of $\alpha$ and
$\beta$ coincide, i.e., if $\alpha = \sum_{i=1}^{\infty}
x_{i}/(b_{1}b_{2}\ldots b_{i})$
and $\beta = \sum_{i=1}^{\infty} y_{i}/(b_{1}b_{2}\ldots b_{i})$, then
$x_{i}=y_{i}$, for all $i =1,2, \ldots ,m$.
\end{lem}

\section{Definitions of a Random Sequence Relative to the Cantor Expansion}

In this section we propose five definitions for random sequences relative to
their
Cantor expansions and we prove that all definitions are mutually equivalent.
We will fix a computable sequence of scales $f$.

We say that the sequence $\x \in X^{f}$ is {\it Cantor--random} if the real
number
$\alpha^{\x}$ is random (in the sense of Algorithmic Information Theory).
e.g., the sequence corresponding to the binary expansion of $\alpha$ is
random.

Next we define the notion of weakly Chaitin--Cantor random sequence. To this
aim we introduce the Cantor self-delimiting Turing machine (shortly, a
machine), which is a
 Turing machine $C$ processing binary strings and producing
elements of $\pref{\XF}$ such that its
program set (domain)
${\it PROG}_C=\{x\in\{0,1\}^* : C(x) \mbox{  halts}\}$
is   a prefix-free set of strings. Sometimes
we will write $C(x) < \infty$ when $C$ halts on $x$ and $C(x) = \infty$ in
the
opposite case.

The {\em program-size complexity} of the string
$w\in\pref{\XF}$ (relative to $C$)
is  defined by $H_C(w)=\min \{|v| :  v \in \Sigma^*, \ C(y)=w\}$,
where $\min \emptyset = \infty$. As in the classical situation
the set of Cantor self-delimiting Turing machines is computably enumerable,
so we can effectively construct  a  machine $U$ (called  {\em universal}\,)
such
that  for every  machine
$C$, $H_U (x) \leq H_C (x) + O(1)$.
In what follows we will fix a universal machine $U$ and denote $H_U$ simply
by $H$.

The sequence $\x \in X^{f}$ is {\it weakly Chaitin-Cantor--random} if there
exists
a positive constant $c$ such that for all $n\in \bbbn$,  $H(\x(n)) \ge
\parallel x \parallel -c$.

The sequence $\x \in X^{f}$ is {\it strongly Chaitin-Cantor--random} if the
following relation holds true:  $\lim_{n
\to \infty} (H(\x(n)) - \parallel x \parallel) = \infty$.

The sequence $\x \in X^{f}$ is {\it Martin-L\" of-Cantor--random} if   for
every computably enumerable collection of computably enumerable open sets
$(O_n)$ in $\XF$ such that for every $n\in \bbbn$,  $\mu (O_n)
\le 2^{-n}$ we have $\x \not\in \cap_{n=1}^{\infty} O_n$.

The sequence $\x \in X^{f}$ is {\it Solovay-Cantor--random} if   for every
computably enumerable collection of computably enumerable open sets $(O_n)$
in $\XF$ such that $\sum_{n=1}^{\infty}\mu (O_n) < \infty$ the relation  $\x
\in  O_n$ is true only for finitely many $n\in\bbbn$.

\begin{theo}
Let $\x \in\XF$. Then, the following statements are equivalent:
\begin{enumerate}
% \item The sequence $\x$ is Cantor--random.
\item The sequence $\x$ is
weakly Chaitin-Cantor--random.
\item The sequence $\x$ is strongly Chaitin-Cantor--random.
\item The sequence $\x$ is
Martin-L\" of-Cantor--random.
\item The sequence $\x$ is Solovay-Cantor--random.
\end{enumerate}
\end{theo}
These equivalences are direct translations of the classical proofs
 (see, for example, \cite{Ca}).

Moreover, we have the following relations.
\begin{theo}
Let $\x \in\XF$. Then, the sequence $\x$ is weakly
Chaitin-Cantor--random if $\x$ is Cantor--random.  If the function $f$
is bounded, then every  weakly
Chaitin-Cantor--random $\x$ is also Cantor--random sequence.
\end{theo}

\begin{proof} \comment{Only the equivalence between the notions of Cantor--randomness
and
weakly Chaitin-Cantor--randomness will be proven% ; all other equivalences are
% direct translations of the classical proofs (see, for example, \cite{Ca}).
}
The argument  is modification of the proof idea of Theorem 3 in
\cite{ludwig1}.

 Assume first that $\x \in\XF$ is not Cantor--random and let
$\alpha = \alpha^{\x}$. Let  $\y = y_{1}y_{2}\ldots$ be the bits of the
binary expansion of $\alpha$. We shall show that $\y$ is not a binary random
sequence.

Fix an integer $m\ge 1$ and consider the rational
\[\alpha (m) = \sum_{i=1}^{m} \frac{x_{i}}{b_{1}b_{2}\ldots
b_{m}}\raisedot\]
We note that $w=x_{1}x_{2}\ldots x_{m}$ is in $\pref{\XF}$
and $\parallel w \parallel = \log (b_{1}b_{2}\ldots b_{m})$. Further on,  $0
< \alpha(m) < \alpha$ and
\[\alpha - \alpha(m)   \le  \sum_{t=m+1}^{\infty}
\frac{x_{t}}{b_{1}b_{2}\ldots b_{t}}
  \le  \sum_{t=m+1}^{\infty} \frac{b_{t}-1}{b_{1}b_{2}\ldots b_{t}}
=  \frac{1}{b_{1}b_{2}\ldots b_{m}}\raisedot
\]
 Next we define the following parameters:
 \begin{equation}
\label{M}
M_{m} = \lfloor \log (b_{1}b_{2}\ldots b_{m})\rfloor,
\end{equation}
\begin{equation}
\label{a}
a_{m} = \lfloor \alpha(m)\cdot 2^{M_{m}}  \rfloor.
\end{equation}
and we note that
\begin{equation}
\label{cantorerror}
 \alpha - \alpha(m) \le \frac{1}{b_{1}b_{2}\ldots b_{m}} \le 2^{-M_{m}}.
 \end{equation}

We are now in a position to  prove the relation: for every integer $m\ge 1$,

\begin{equation}
\label{inclusion}
[\alpha(m), \alpha] \subseteq \left[a_{m}\cdot 2^{-M_{m}}, (a_{m}+2)\cdot
2^{-M_{m}}\right).
\end{equation}
Indeed, in view of (\ref{cantorerror})
and (\ref{a}) we have $\alpha < (a_{m}+2)\cdot 2^{-M_{m}}$ as:
\[\alpha \cdot 2^{-M_{m}} \le \alpha(m) \cdot 2^{-M_{m}} +1 < a_{m} +2. \]
Again from (\ref{a}), $ a_{m} \le \alpha(m)\cdot 2^{M_{m}}$.

Using (\ref{inclusion}),  from $w= x_1 x_2 \ldots x_m$ plus two more bits we
can determine $y_{1} y_{2} \ldots y_{M_{m}}$, that is, from the first $m$
digits of the Cantor expansion of $\alpha$ and two additional bits we can
compute the first $M_{m}$ binary digits of $\alpha$. In view of
Lemma~\ref{binapprox} we obtain
a computable function $h$ which  on  an input consisting of a binary string
$v$ of length 2
and $w$  produces as output $\y (M_{m})$.

We are ready to use the assumption  that
$\y$ is  random but $\x$ is not Cantor--random, that is, there is
  a universal self-delimiting Turing machine $U^{2}$ working
on binary strings and there is
 a positive constant $c$ such that for all $n\ge 1$,
\begin{equation}
\label{br}
H_{U^{2}} (\y(n)) \ge n-c,
\end{equation}
and
 for every positive $d$ there exists a positive integer $l_{d}$ (depending
upon $d$) such that
\begin{equation}
\label{not-cr}
H(\x(l_{d})) \le \, \parallel \x(l_{d})
\parallel - \, d.
\end{equation}
We construct a binary self-delimiting Turing machine $C^{2}$
such that for every $d>0$, there exist two strings $l_{d}$ and $v,
s_{l_{d}}\in \{0,1\}^{*}$, such that $|v| = 2,$  $|s_{l_{d}}| \le
\,\parallel  \x(l_{d}) \parallel -d = \log (b_{1} b_{2}\ldots b_{l_{d}}) -\, d$
and $C^{2}(v, s_{l_{d}}) = \y(M_{l_{d}})$.

Consequently, in view of (\ref{br}) and (\ref{not-cr}),  for every $d$ we
have:
\begin{eqnarray*}
M_{l_{d}} -c & \le &  H_{U^{2}}(\y(M_{l_{d}}))\\
& \le &  H_{C^{2}}(\y(M_{l_{d}})) + O(1)\\
& \le & |s_{l_{d}}| + 2 + O(1)\\
& \le & \log (b_{1} b_{2}\ldots b_{l_{d}}) + O(1)\\
& = &  M_{l_{d}} + O(1) -  \,d,\\
\end{eqnarray*}
a contradiction.

Recall that $\alpha = \sum_{i=1}^{\infty} x_{i}/(b_{1} b_{2}\ldots b_{i}) =
\sum_{i=1}^{\infty} y_{i}2^{-i}$. Now we prove that $\x$ is Cantor--random
whenever $\y$ is random. Let $m\ge 1$ be an integer and let $\alpha_{2}(m)
= \sum_{i=1}^{m} y_{i}2^{-i}$. Given  a large enough $m$ we effectively
compute the integer $t_{m}$ to be the maximum integer $L\ge 1$ such that
\begin{equation}
\label{tm}
2^{-m} \le \frac{1}{b_{1}b_{2}\ldots b_{L}}\raisedot
\end{equation}
We continue by proving that for all large enough  $m\ge 1$:
\begin{equation}
\label{cantorineq}
[\alpha_{2}(m), \alpha] \subseteq \left[\alpha (t_{m})
-
\frac{1}{b_{1}b_{2}\ldots b_{t_{m}}}, \alpha (t_{m}) +
\frac{1}{b_{1}b_{2}\ldots b_{t_{m}}}\right]\raisedot
\end{equation}
We note that $\alpha_{2}(m) < \alpha$ and

\[\alpha  =   \sum_{i=1}^{\infty} \frac{x_{i}}{b_{1}b_{2}\ldots b_{i}}
 \le   \alpha (t_{m}) + \sum_{i=t_{l}}^{\infty}
\frac{x_{i}}{b_{1}b_{2}\ldots b_{i}}
 \le  \alpha (t_{m}) + \sum_{i=t_{l}}^{\infty}
\frac{b_{i}-1}{b_{1}b_{2}\ldots b_{i}} \\
 \le  \alpha (t_{m}) + \frac{1}{b_{1}b_{2}\ldots b_{t_{m}}} \raisedot
\]

\if01
\begin{eqnarray*}
\alpha & = &  \sum_{i=1}^{\infty} \frac{x_{i}}{b_{1}b_{2}\ldots b_{i}} \\
& \le &  \alpha (t_{m}) + \sum_{i=t_{l}}^{\infty}
\frac{x_{i}}{b_{1}b_{2}\ldots b_{i}} \\
& \le & \alpha (t_{m}) + \sum_{i=t_{l}}^{\infty}
\frac{b_{i}-1}{b_{1}b_{2}\ldots b_{i}} \\
& \le & \alpha (t_{m}) + \frac{1}{b_{1}b_{2}\ldots b_{t_{m}}} \raisedot
\end{eqnarray*}
\fi

As $\alpha \le \alpha (t_{m}) + 1/(b_{1}b_{2}\ldots b_{t_{m}})$ we only need
to show that $\alpha (t_{m}) \le \alpha_{2} (m)+ 1/(b_{1}b_{2}\ldots
b_{t_{m}})$. This is the case as otherwise, by  (\ref{tm}), we would have:

\[\alpha (t_{m}) > \alpha_{2} (m) +
 \frac{1}{b_{1}b_{2}\ldots b_{t_{m}}} \ge \alpha_{2} (m) + 2^{-m}
 \ge \alpha,\]
 a contradiction.

In case when $f$ is bounded, assume by contradiction that $\x$ is
Cantor--random but $\y$ is not random, that is there exists a positive
constant $c$ such that for all $n\ge 1$ we have:
\begin{equation}
\label{crand}
H(\x(n)) \ge \log (b_{1}b_{2}\ldots b_{n}) -c,
\end{equation}
and for every $d>0$ there exists an integer $n_{d}>0$ such that
\begin{equation}
\label{not-brand}
H_{U^{2}} (\y(n_{d})) < n_{d} - d.
\end{equation}
In view of Lemma~\ref{cantorapprox} and (\ref{cantorineq}) there is a
computable function $F$ depending upon two binary strings such that $|v|=2$,
$F(\y (n_{d}), v) =
\x (t_{n_{d}})$, so the partially computable function $F \circ U^{2}$
which
maps binary strings in elements of $\pref{\XF}$ is a
Cantor self-delimiting
Turing machine such that for every $d>0$ there exists a binary string
$s_{n_{d}}$ of length less than
$n_{d} - d$ and a binary string $v$ of length  2 such that
$F(U^{2}(s_{n_{d}}), v) = \x(t_{n_{d}})$.

As $f$ is bounded, the difference  $\mid t_{m+1}-  t_{m}\mid$ is bounded.  In view of (\ref{tm}), for large $m\ge 1$,
$b_{1}b_{2}\ldots b_{t_{m}} > m-1$, so we can write:
\if01
\begin{equation}
\label{errorcb}
2^{-n_{d}} + \sum_{i=1}^{n_{d}} (1-y_{i})2^{-i} = 1 - \sum_{i=1}^{n_{d}}
y_{i}2^{-i} \le  1 - \frac{1}{b_{1}b_{2}\ldots b_{t_{n_{d}}}}\raisecomma
\end{equation}
\fi
\begin{eqnarray*}
n_{d} - c -1  & \le & \log (b_{1}b_{2}\ldots b_{t_{n_{d}}}) - c\\
 & \le &  H_{U}(\x (t_{n_{d}}))\\
 & \le & H_{F \circ U^{2}} (\x (t_{n_{d}})) + O(1)\\
 & \le & |s_{n_{d}}| +  2 + O(1)\\
& \le & n_{d} -d +  O(1),
\end{eqnarray*}
a contradiction.
\end{proof}
%% Remark: I didi not change the proof, but we should possibly include a
%% remark where it is important to have $f$ bounded.

\begin{open} It is an open question whether the above result holds true for unbounded
functions $f$.
\end{open}

Consider the following statement:

 \begin{center}
 \it
 Let $\x$ be a binary sequence. If there exists a computable infinite set
 $M$ of positive integers and $c>0$ such that for every $m\in M$, $H_{U^{2}}(\x(m))
 \ge m-c$, then $\x$ is random.
 \end{center}

 Note that if the above statement would be true,
then the answer to the Open Question would be affirmative.

It is interesting to note that in case of unbounded functions $f$ we
may have Cantor--random sequences $\x\in \XF$ which do not contain a
certain letter, e.g. $0\in X_i$.

\begin{ex}
Let $f(i)= 2^{i+2}$. Then the measure of the set $F=
\prod_{i=1}^\infty X'_i$, where $X'_i=X_i\setminus \{0\}$ satisfies
$\mu(F) = \prod_{i=1}^\infty (1- 2^{-i-1})>0$. Thus $F$ contains a
Cantor--random sequence $\x$.

However, by construction, $\x$ does not contain the letter $0$ which is in every $X_i$.
\end{ex}

\section{On the Meaning of Randomness in Cantor's Setting}

So far, a great number of investigations have concentrated on the meaning
and definition of randomness in the standard context, in which bases remain
the same at all scales. That is, if one for instance ``zooms into''  a
number by considering the next place in its expansion, it is always  taken
for granted that the same base is associated with different places.

 From a physical viewpoint, if one looks into a physical property encoded
into a real in, say, fixed decimal notation,  then by taking the next digit
amounts to specifying that physical property more precisely by a factor of
ten. A fixed ``zoom'' factor may be the right choice if all physical
properties such as  forces and symmetries and boundary conditions remain
the same at all scales. But this is hardly to be expected. Take, for
instance, a ``fractal'' coastline. How is it generated? The origins of its
geometry are the forces of the tidal and other forces on the land and
coastal soil. That is, water moving back and forth, forming eddies, washing
out little bays, and little bays within little bays, and little bays within
little bays within little bays, \ldots \, and so on. There may be some structural
components of this flow which results in scale dependence. Maybe the
soil-water system forming the landscape will be ``softer'' at smaller
scales, making bays relatively larger that their macroscopic
counterparts.  Indeed, eventually, at least at subatomic scales, the
formation of currents and eddies responsible for the creation of ever
smaller bays will break down.

In such cases, the base of the expansion might have to be modified in order
to be able to maintain a proper relation between the coding of the
geometric object formed by the physical system and the meaning of its number
representation in terms of ``zooming''.  All such processes are naturally
stochastic, and therefore deserve a proper and precise formalization in
terms of random sequences in Cantor representations.

%\bibliography{svozil}

\begin{thebibliography}{99}%
\bibitem{Ca} C.~S.~Calude. \textsl{Information and Randomness. An
Algorithmic
    Perspective}, 2nd Edition, Revised and Extended, Springer Verlag,
Berlin,
  2002.

  \bibitem{ch8}
G. J. Chaitin\index{Chaitin, G.J.}. {\em Algorithmic Information Theory},
Cambridge University Press, Cambridge,  third printing 1990.

\bibitem{gregc2}  G. J. Chaitin.\index{Chaitin, G.J.} {\em Exploring
Randomness}, Springer-Verlag,
London, 2001.

\bibitem{drobot} S. Drobot. {\em Real Numbers}, Prentice-Hall, Englewood
Cliffs, New Jersey, 1964.

\bibitem{hw}
G. H. Hardy\index{Hardy, G.H.}, E. M. Wright\index{Wright, E.M.}. {\em An
Introduction to the Theory of
Numbers}, Clarendon Press, Oxford, 5th ed., 1979.

\bibitem{mar1} P. Martin-L\"{o}f. The definition of random sequences,
  \textsl{Inform.  and Control} 9 (1966), 602--619.

\bibitem{mar2} P. Martin-L\"{o}f. \textsl{Notes on Constructive
Mathematics},
  Almqvist \& Wiksell, Stockholm, 1970.

\bibitem{sum-4}
J. Summhammer.
 The physical quantities in the random data of neutron interferometry,
in E.~I. Bitsakis,  C.~A. Nicolaides, eds, {\em The Concept of
  Probability},  Kluwer Akademic Publishers, Amsterdam, 1989.


\bibitem{ludwig1}  L. Staiger. The Kolmogorov complexity of real numbers,
 {\em Theoret. Comput. Sci.} 284 (2002), 455--466.

\bibitem{WooFie}
W. K. Wootters,  B.D. Fields.
Optimal state-determination by mutually unbiased measurements,
 {\em Annals of Physics} 191 (1989), 363--381.



\end{thebibliography}
%\bibliographystyle{apsrev}
%\bibliographystyle{unsrt}

\end{document}